\documentclass[%
 jmp,
 amsmath,amssymb,
twocolumn,%
]{revtex4-1}

\usepackage{graphicx}% Include figure files
\usepackage{dcolumn}% Align table columns on decimal point
\usepackage{bm}% bold math
\usepackage[allcolors=blue,colorlinks=true]{hyperref}
\usepackage[utf8]{inputenc}
\usepackage[T1]{fontenc}
\usepackage{mathptmx}
\usepackage{etoolbox}
\usepackage{diagbox}
\makeatletter
\def\@email#1#2{%
 \endgroup
 \patchcmd{\titleblock@produce}
  {\frontmatter@RRAPformat}
  {\frontmatter@RRAPformat{\produce@RRAP{*#1\href{mailto:#2}{#2}}}\frontmatter@RRAPformat}
  {}{}
}%
\makeatother

\begin{document}

\begin{abstract}

Rigid bodies, made of smaller composite beads, are commonly used to simulate anisotropic particles with molecular dynamics or Monte Carlo methods. To accurately represent the particle shape and to obtain smooth and realistic effective pair interactions between two rigid bodies, each body may need to contain hundreds of spherical beads. Given an interacting pair of particles, traditional molecular dynamics methods calculate all the inter-body distances between the beads of the rigid bodies within a certain distance. For a system containing many anisotropic particles, these distance calculations are computationally costly and limit the attainable system size and simulation time. However, the effective interaction between two rigid particles should only depend on the distance between their center of masses and their relative orientation. Therefore, a function capable of directly mapping the center of mass distance and orientation to the interaction energy between the two rigid bodies, would completely bypass inter-bead distance calculations. It is challenging to derive such a general function analytically for almost any non-spherical rigid body. In this study, we have trained neural nets, powerful tools to fit nonlinear functions to complex datasets, to achieve this task. The pair configuration (center of mass distance and relative orientation) is taken as input and the energy, forces and torques between two rigid particles are predicted directly. We show that molecular dynamics simulations of cubes and cylinders performed with forces and torques obtained from the gradients of the energy neural-nets quantitatively match traditional simulations that uses composite rigid bodies. Both structural quantities and dynamic measures are in agreement, while achieving up to 23 times speed up over traditional molecular dynamics, depending on hardware and system size. The method presented here can, in principle, be applied to any irregular shape with any pair interaction, provided that sufficient training data can be obtained. 
\end{abstract}

\title{Molecular Dynamics Simulations of Anisotropic Particles Accelerated by Neural-Net Predicted Interactions}

\author{B. Ru\c{s}en Argun}
\affiliation{Mechanical Engineering, 
                              Grainger College of Engineering, University of Illinois,Urbana-Champaign, 61801, IL, United States}

\author{Yu Fu}
\affiliation{Physics, 
                              Grainger College of Engineering, University of Illinois,Urbana-Champaign, 61801, IL, United States}
                
\author{Antonia Statt}
\email{statt@illinois.edu}
\affiliation{Materials Science and Engineering, 
                              Grainger College of Engineering, University of Illinois, Urbana-Champaign, 61801, IL, United States}

\maketitle

\section{Introduction}
 Advances in chemistry have made it possible to synthesize many non-spherical colloidal particles \cite{sacanna2013shaping}. 
 Those anisotropic colloidal building blocks can self-assemble into various targeted nano-structures with desirable structural, optical or electronic properties \cite{cai2021colloidal,henzie2012self}. 
In the environment, pollutants such as microplastics or engineered nano-particles can occur in different shapes \cite{argun2023influence, wang2021review}. 
In addition, particles with non-spherical morphologies are central to several biological processes \cite{roth1996van}. 
Thus, capturing anisotropic features of particles is essential to accurately simulate and understand the fundamentals of wide range of phenomena in science and engineering. 

Traditional particle based simulations such as molecular dynamics (MD) and Monte Carlo (MC) simulations often treat particles as perfect spheres, however, there are several methods available that can take non-spherical geometric features of particles into account. 
%%--- Hard Core repulsion methods --- 
For athermal systems, where particles interact with hard repulsion only, calculating the pairwise potential reduces to the task of finding whether the two particles or shapes overlap or not. 
In such systems, where the self-assembly is driven by entropic effects, powerful methods to calculate the interactions for many geometries exist. 
\citet{donev2006jammed} introduced an event-driven MD algorithm that can efficiently calculate the overlap between non-spherical geometries. 
This method has been used to find closed-packing structures and phase behaviour of hard superballs \cite{ni2012phase,donev2004improving}.
\citet{damasceno2012predictive} and others\cite{anderson2016scalable,geng2019engineering,van2015digital} have simulated a wide range of polyhedra with hard-particle MC and investigated their self-assembly and phase behaviour using different methods to calculate the overlap or contact between the shapes \cite{gilbert1988fast}. 
%%--- Attraction Methods ---- 
However, for systems where the self-assembly is driven by enthalpic forces (i.e. attractive interactions), different methods are required. 
The Gay-Berne \cite{berardi1998gay} potential is commonly used to simulate ellipsoids with pairwise attractive interactions, depending on the orientations of the particles in addition to the distance between their center of masses. 
It is convenient for MD simulations, as forces and torques can be calculated analytically, however, it is limited to ellipsoid shapes only. The anisotropic Lennard-Jones potential~\cite{ramasubramani2020mean} extends this to  polyhedra. 

By using a composite bead approach, \textit{any} particle shape with attractive and hard repulsive interactions can be simulated. Due to its flexibility in both shape and interaction, this method is commonly used. 
Here, smaller, spherical beads, often called composite beads, are held together with stiff harmonic bonds and angle potentials to keep the geometry of the larger composite shape the same throughout the simulation. Alternatively, rigid body constraints between the composite beads can be applied. 
Popular MD software such as HOOMD-blue~\cite{anderson2020hoomd,nguyen2011rigid,glaser2020pressure} and LAMMPS~\cite{plimpton1995fast} have implemented rigid body constraints to simulate non-spherical particles or molecules made out of smaller beads \cite{nguyen2011rigid, nguyen2019aspherical}.  
The composite bead approach is flexible, as almost any concave or convex shape can be simulated with it. 
Self-assembly behaviour of nanoparticles can be affected by small morphological details \cite{kim2017imaging}. 
To obtain a precise geometry including sharp corners and edges, the shape often needs to contain many beads. 
The interaction between two bodies or shapes is now given by the double sum of all pairwise interactions between the composite beads that make up the two bodies. Thus, rigid bodies containing many composite beads significantly increase the computational cost in both coarse-grained and atomistic simulations. 
A similar computational cost challenge arises for example to calculate the total van der Waals (vdW) interaction between two non-spherical particles.
Most notably, \citet{lee2020analytical} devised an analytical method to approximate the vdW interactions between two cubic rigid bodies bypassing the cumbersome double sum.
Their method allows for MC simulations of cubes, but it is limited in its application to MD since forces and torques cannot be obtained with it. 
There are several similar efforts in the literature to calculate the vdW interaction for non-spherical bodies \cite{priye2013computations, jaiswal2012approximate, hallez2012analytical, yang2009general, maeda2015orientation}. 
Recent work \cite{lee2023effects} has calculated vdW forces and torques for cubes with a numerical method.  
\citet{ramasubramani2020mean} introduced a Lennard Jones-like potential that can be used for different polyhedra. 

Machine Learning (ML) techniques are already applied to spherically symmetric particles, especially to bridge quantum mechanical calculations and classical atomistic MD simulations \cite{deringer2019machine}. 
Some possibilities that ML offers for anisotropic interactions are recently discovered. 
\citet{campos2022machine} has developed symmetry functions that can be used to construct ML potentials for cylindrically symmetric particles.  
Anisotropic and rigid molecules like benzene were coarse-grained by matching forces and torques to accelerate all-atom condensed-phase MD simulations using a linear model \cite{nguyen2022systematic} and neural-networks \cite{wilson2023anisotropic}.    
Recent work \cite{bag2023machine} is using several ML models to detect whether or not two composite rigid bodies are overlapping, bypassing explicit distance calculations between primary beads of the rigid bodies. However, this method does not allow attractive interactions.  

In this work, we are using fully-connected feed-forward neural nets to predict the total interaction between two example geometries of composite rigid bodies: cubes and cylinders. We focus on forces and torques, since these are the necessary inputs for a MD simulation.
Here, we describe our generic method of data sampling and processing and test the neural-net predictions for simple example particle shapes.  
We show that the MD simulations performed with forces and torques from the trained neural nets accurately reproduce structural and dynamic properties of traditional rigid body simulations which were performed with explicit distance calculations.
Depending on the system simulated (number of particles, temperature etc.) and hardware available, neural-net assisted simulations are faster than traditional MD simulations. Our method is entirely agnostic to the particle shape and can be, in principle, applied to any irregular composite particle geometry with a wide range of both attractive and repulsive pair interactions.

\section{Methods and Motivation}
\begin{figure}
    \centering
    \includegraphics[width=8cm]{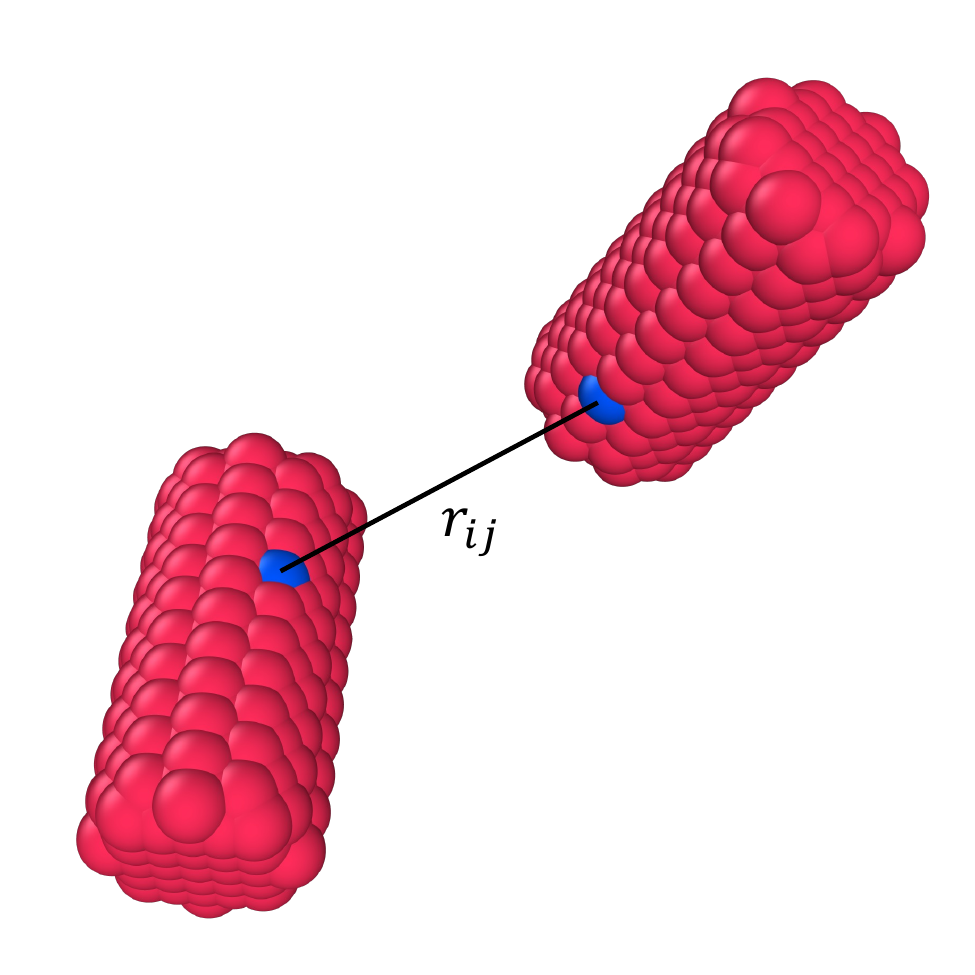}
    \caption{A pair of cylindrical rigid-bodies made out of smaller beads. The cylinder is approximately $2.75\sigma$ in diameter and $5.3\sigma$ long.}
    \label{fig:the_pair}
\end{figure}

To calculate the total potential $U$ between a pair of composite-rigid bodies that are made out of smaller beads (as in Fig. \ref{fig:the_pair})  that interact with pairwise potentials $u_{bead}(r)$, the distances between each pair of beads of need to be calculated. 
\begin{equation*}
    U = \sum_{i=1}^{N} \sum_{j=1}^{N} u_{bead}(r_{ij}) \quad,
\end{equation*}
For two rigid bodies, each consisting of $N$ beads, calculating $N^2$ distances $r_{ij}$, can be computationally costly for high values of $N$.
The number of beads $N$ used to represent a rigid-body depends on several factors, including the geometry of the particle and how accurate one would like to reproduce that geometry. 
For example, to perform a systematic study of particle shapes that range from a perfect cube to a sphere with several super-ball shapes in between, edges and corners of the cube must be smoothed out gradually as the shape approaches to a sphere, requiring a high resolution of beads.
Simulation methods that are used to incorporate a background fluid and hydrodynamic interactions like multi-particle collision dynamics \cite{malevanets1999mesoscopic} may also require a high resolution of shape surface for accurate coupling between the colloid and solvent particles \cite{poblete2014hydrodynamics,wani2022diffusion}.
Another factor is the type of interaction between the shapes. 
For hard-core repulsion interactions or WCA-like \cite{WCA} interactions, where there is no attraction, one can place beads at the surface of the shape only and leave the core empty, thereby reducing $N$ and saving computational time. 
However, this approach may not work for attractive interactions.
By only having surface beads, or shapes with a small number of beads, it may not be possible to reproduce the desired effective interaction between the two shapes in such cases. $N$ may need to be on the order of $10^2$ to accurately represent the shape and reproduce the desired effective interaction. 
We refer to the SI for a detailed illustration of this issue. 

In the following, we outline an alternative route to reduce computational costs associated with the composite bead method by bypassing explicit distance calculations, while keeping the full flexibility of the different shapes that can be used.
Any pair configuration of $3D$ bodies can be expressed as a function of six numbers.
Position and orientation of the first body 1 define a coordinate frame.
The position of the second body 2 $\mathbf{x}(x_{1},x_{2},x_{3})$ and its orientation  $\boldsymbol{\Omega}(\Omega_{1},\Omega_{2},\Omega_{3})$ in the relative coordinate frame of the first body 1 are then sufficient to uniquely define any pair configuration.
Thus, the total interaction between the two composite-rigid bodies is essentially a function of six numbers,
\begin{equation*}
    U = g(\mathbf{x},\boldsymbol{\Omega}) = g(x_{1},x_{2},x_{3},\Omega_{1},\Omega_{2},\Omega_{3}) \quad.
\end{equation*}
However, this function is quite complex and its general analytical solution is not available even for simple shapes like cubes or non-aligned cylinders \cite{lee2020analytical}. 
We follow a data-driven approach and use neural-nets ($g_{NN}$) to approximate the function $g$ relating the pair configuration directly to the interaction energy, force or torque, 
\begin{equation*}
    U = \sum_{i=1}^{N} \sum_{j=1}^{N} u_{bead}(r_{ij})= g(\mathbf{x},\boldsymbol{\Omega}) \approx g_{NN_{U}}(\mathbf{x},\boldsymbol{\Omega}) \quad,
\end{equation*}
thus allowing us to skip the costly distance calculations at the expense of the inference and training cost of a neural-net.

Training and testing data for the neural net was sampled from traditional composite-rigid body MD simulations that were performed at a wide range of temperatures, starting from $T= 0.2 \epsilon/k_B$ up to $2.0 \epsilon/k_B$. 
Both energetically favorable (i.e., pairs that are strongly attracting, for example two cubes with their faces aligned) and unfavorable (i.e., pairs that are slightly overlapping or touching, physically less likely to occur) pair configurations can be efficiently sampled at low and high temperatures, respectively. 
Sampled body pairs and their calculated energy, force and torque vectors are then subjected to  geometric transformations and operations in order to express each pair as a function of $\boldsymbol{\Omega}$ and $\mathbf{x}$ and to make use of the geometrical symmetries of the shapes.
This process is explained in detail in the SI Section 1 and the code is publicly available at \url {https://github.com/stattlab/NNAssistedRigidBodyMD}. 
Here, we only mention that unlike  other $3D$ bodies, a pair of cylinders can be expressed as a function of four numbers only $g(\mathbf{x}(x_1,x_2), \boldsymbol{\Omega}(\Omega_1,\Omega_2) )$, thanks to their cylindrical symmetry. 

Training a single neural-net that can predict the energy would be sufficient to perform Monte Carlo simulations. 
For MD simulations, force $\mathbf{f}=(f_1,f_2,f_3)$ and torque $\boldsymbol{\tau}=(\tau_1,\tau_2,\tau_3)$ vectors are required to integrate the equations of motion forward in time. 
We have used two different methods to obtain $\mathbf{f}$ and $\boldsymbol{\tau}$. 
First, for cylinders, five separate neural nets were trained for each component of $\mathbf{f}=(f_1,f_2,f_3)$ and $(\tau_1,\tau_2)$ and used directly in the simulation. The torque around the axis of the cylinder, $\tau_{3}$ is not needed.
This leads to the following approximations
\begin{equation*}
    f_{i} \approx g_{NN_{f_{i}}}, \quad \tau_{i} \approx g_{NN_{\tau_{i}}} \quad \text{with} \quad i = {1,2,3} \quad.
\end{equation*}
Second, for cubes of side length approx. $3.2\sigma$, the forces and torques were calculated from the gradients of the energy neural-net as explained in SI Section 6.
The second method can provide the same accuracy as the first one and it is more practical to implement. In addition, using one neural net for energy also helps to ensure that the components of the forces and torques obey energy conversation. 
Simple fully connected feed-forward neural nets are used for these regression tasks.

\section{Results and Discussion} 

\begin{figure}
    \centering
    \includegraphics[width=8cm]{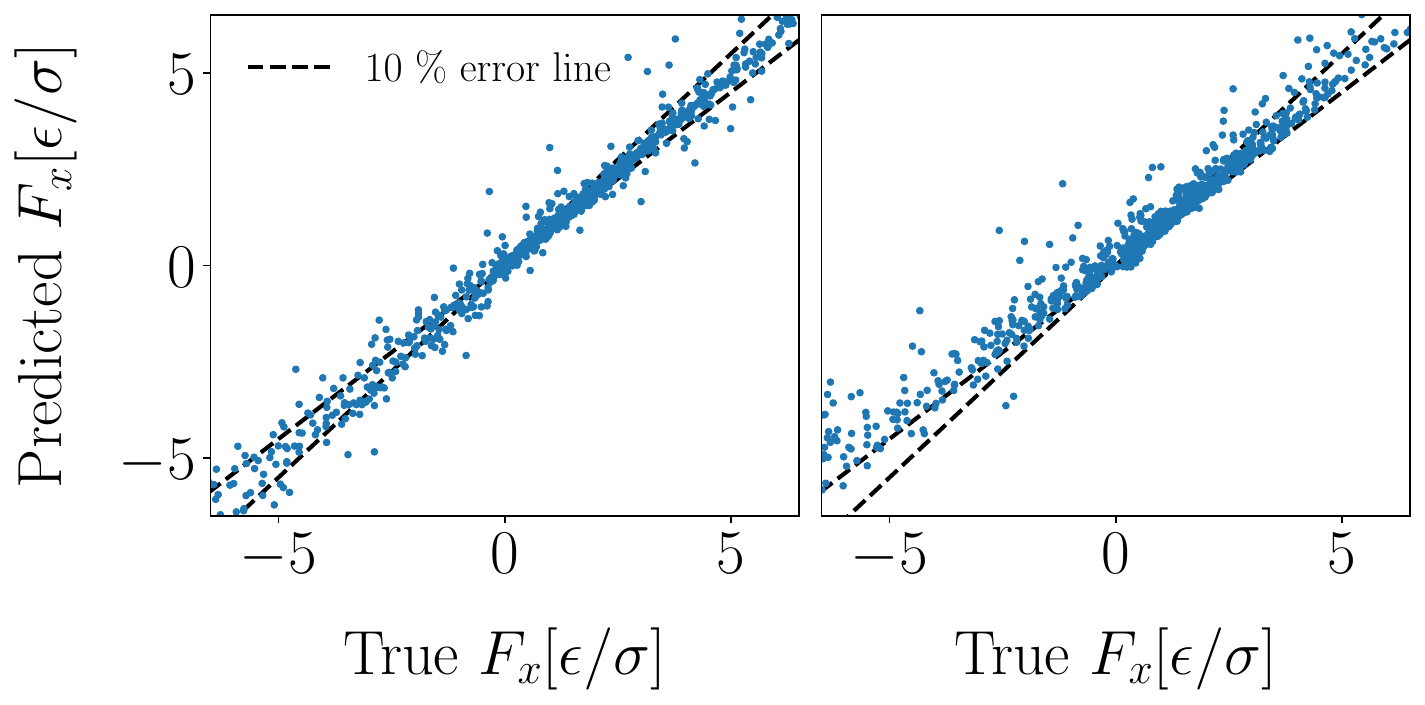}
    \caption{Parity plot of true vs. predicted $f_{x}$ values for the cylinder (left) and cube (right) system. For cylinder $f_{x}$ is the direct output of the force neural net $g_{NN_{f_{i}}}$, for cube it is obtained from the gradient of the energy neural net $g_{NN_{U}}$. Dashed lines indicate the 10\% error line.}
    \label{fig:parity}
\end{figure}

Generally, accurate force-torque prediction from our model should result in correct molecular dynamics simulations. Before comparing the traditional and neural-net assisted simulation results, we are presenting a more direct measure of ML accuracy, namely parity plots obtained from the test data. 
In Fig.~\ref{fig:parity}, the predicted values of the $x$-component of the net force ($f_{x}$) vs. the true value between two cylinders and cubes are plotted. The other force components follow the same trends. 
%Most of the 1500 data points for the cubes were inside the 10\% error lines,  as the force values get closer to zero the relative error rate inevitably increased.  
For both shapes, data points outside the 10\% error lines were frequent, especially for values close to zero.
We note that using separate neural networks to predict $f_{i}$ for cubes, instead of taking the gradient of the $g_{NN_{U}}$ results in a much lower rate of error, as we show in SI Section 5. 
However, since the accuracy of the MD simulation results are similar for both methods, the results obtained from cube simulations assisted by $g_{NN_{f_{i}}}$ and $g_{NN_{\tau_{i}}}$ are not presented here.  
% Since the data contained many values close to zero, a mean absolute percentage error or other common relative errors that have the true value in the denominator are of limited use. 
% Here, we chose to simply report the rate of data points outside of the 10\% error lines, which is 32\% for cylinders and 11\% for cubes.

\begin{figure}
    \centering
    \includegraphics[width=8cm]{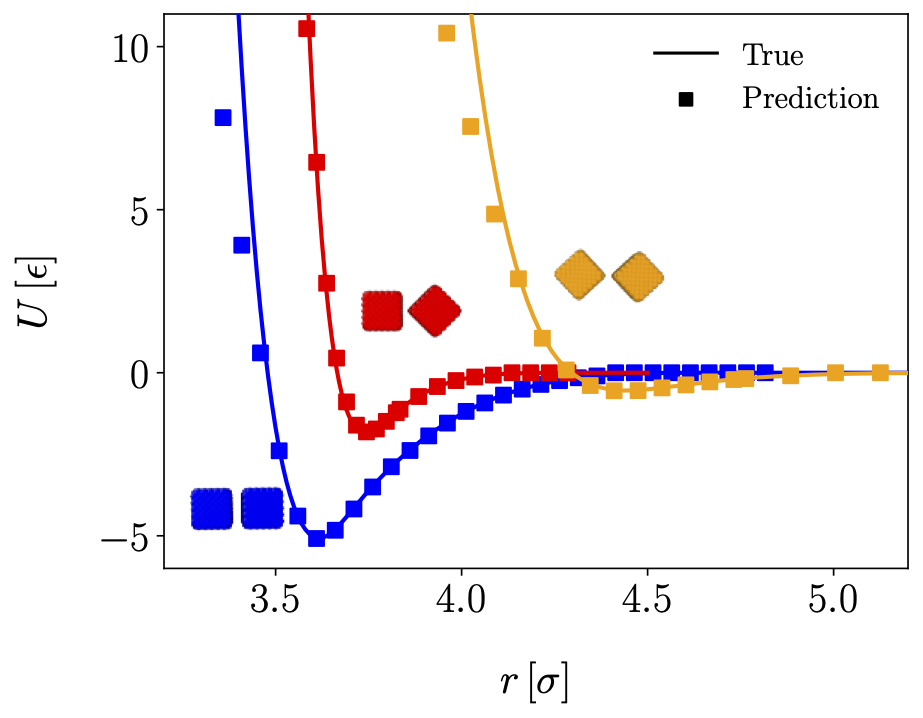}
    \caption{Interaction energy $U(r)$ as function of center of mass distance $r$ for three different cube configurations, as shown next to the curves. Each cube had an approximate side length of $3.2\sigma$.}
    \label{fig:energy_vs_dist}
\end{figure}

In Fig.~\ref{fig:energy_vs_dist}, traditionally calculated and neural-net predicted energies of cubes are compared. Here, we show the energies of three different relative orientations as a function of their center of mass distances. 
Overall, there is good agreement over a wide range of center of mass distances, with noticeable deviations for small distances and high interaction energies. These short distances are less likely to occur in a molecular dynamics simulation, but might play a significant role for simulations at highly dense conditions or higher temperatures above the temperatures used to generate the training data set. 

The next step is to compare results of neural-net assisted simulations with traditional molecular dynamics simulations.  
For this, we performed two sets of simulations, one using HOOMD-blue 2.9.4 with traditional molecular dynamics simulations of rigid bodies in the NVT ensemble, and the second set with C\texttt{++} code using neural-net forces and torques for the molecular dynamics simulation.
We are using half-power half-cosine as a base function, 
\begin{equation}
\mathrm{U}(r)= \begin{cases}-A \cos (c(r-b)) & r>b \\ (b-r)^p-2 A & r \leq b\end{cases}\quad,
\end{equation}
for the bead-to-bead potential as detailed in the SI Section 4 with the following parameters, $A=0.00035\epsilon$, $b=1.46375\sigma$, $c=3.3833\sigma^{-1}$, $p=2.5$ for both cylinders and cubes.
When summed up over the beads of the two interacting cubes, we effectively obtain a Lennard Jones-like interaction between the two cubes as illustrated in Fig. ~\ref{fig:energy_vs_dist}.

The temperature was kept constant using a Nosé–Hoover thermostat with a timestep of 0.005 $\tau$. We equilibrated the system for $2500 \tau$, where $\tau = \sigma\sqrt{m/\epsilon}$ is the unit of time. The particle volume fraction $\phi = N*V_\mathrm{particle}/V$ of all simulations was set to $\phi=0.16$.  We used the size of the smaller composite beads $\sigma$, as unit of length.  
Cubes have an approximate side length of $3.2\sigma$ and the cylinders are approx. $2.75\sigma$ in diameter and $5.3\sigma$ long.

\begin{figure}
    \centering
    \includegraphics[width=8cm]{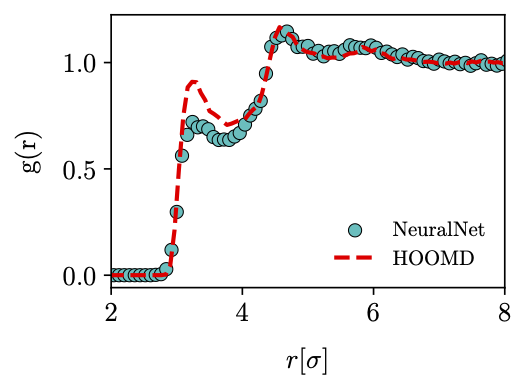}
    \caption{Pair correlation function $g(r)$ at $kT=1.0$ for a cylinder system. The traditional simulation result is shown as dashed line, and the nn-assisted simulation result is shown as filled symbols.}
    \label{fig:pcf_cylinder}
\end{figure}

To assess the accuracy of our simulations a measure for the structure of the system was used, the pair correlation function $g(r)$ of the center of mass of the simulated shapes. 
In Fig.~\ref{fig:pcf_cylinder} $g(r)$ of cylinders was plotted for both nn-assisted and traditional MD simulations. 
While qualitative agreement between the two curves can be observed, and nn-assisted simulations reproduced the peaks at correct positions, the height of the first peak is underestimated. 
The first peak in $g(r)$ corresponds to the pair configuration where two cylinders are aligned along their main axis and are close to each other. 
We attribute the mismatch at the first peak to the assumption in nn-assisted simulations that the cylinders are symmetric around their main axis. 
Even though the cylinders were comprised of many composite beads for the traditional MD simulations, the surface was still rough and rotations around the main axis might at close distances result in different interactions. This roughness is effectively coarse-grained out in the neural-net assisted simulations, since the neural net cannot account for rotations around the main axis by design of the input variables. 
This effect is the strongest when two cylinders are closest to each other, almost touching, and decays as the distance increased, as shown in Fig. SI-S6. 

\begin{figure}
    \centering
    \includegraphics[width=8cm]{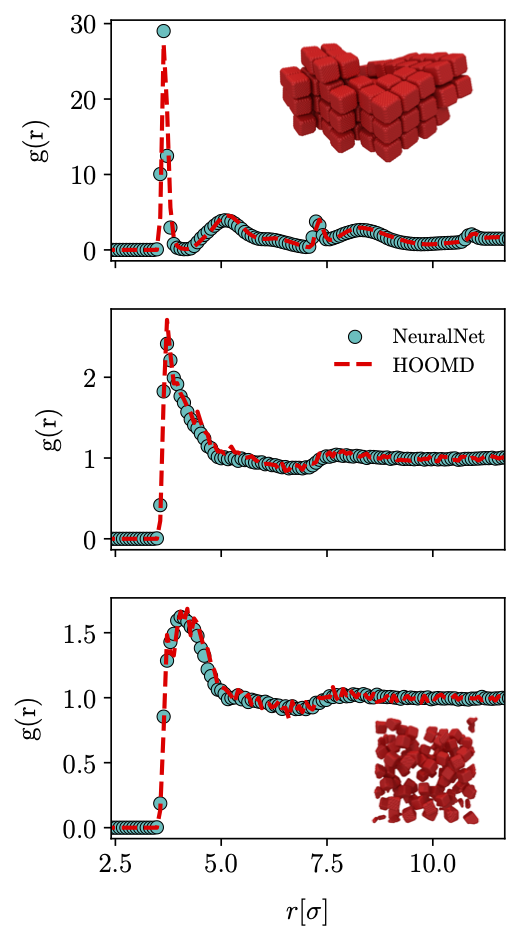}
    \caption{Pair correlation functions $g(r)$ for three different temperatures: $0.5\,\epsilon/k_B$ (top), $0.6\,\epsilon/k_B$ (middle) and $0.7\,\epsilon/k_B$ (bottom) for cubes. Dashed lines show the result from traditional simulations, filled symbols show the result of the nn-assisted simulations. The insets are representative snapshots of the corresponding structures at these temperatures.}
    \label{fig:pcf_03_cube}
\end{figure}

Next, we tested the structural accuracy of the cube simulations. In Fig. \ref{fig:pcf_03_cube}, the $g(r)$ curves for the center of mass of cubes are plotted for simulations performed at temperatures of $T=0.5\,\epsilon/k_B$, $0.6\,\epsilon/k_B$, and $0.7\,\epsilon/k_B$. 
There is quantitative match at all temperatures simulated. 
The nn-assisted simulations accurately capture the significant structural changes observed in this temperature window, ranging from a non-aggregated suspension of cubes at high temperatures $T=0.7\,\epsilon/k_B$, down to a aggregated face-centered cubic structure at low temperatures $T=0.5\,\epsilon/k_B$. 
Unlike in the case of cylinders, where we ignore rotations around the main axis, rotations in the cube configurations are represented explicitly and we speculate that this is part of the reason for the improved match of the first peak in $g(r)$. 

%low ($k_BT=0.3$) and high ($k_BT=0.75$) temperatures, except for the first peak at low temperature which is under-predicted in the nn-assisted simulation. 
%However, at medium temperatures ($k_BT=0.5$), there is a complete mismatch, the neural-net assisted and traditional md simulations pr  oduce two completely different structures. 
%Traditional molecular dynamics simulation seems to be well-ordered/structured with a very strong first peak and clear $2^\text{nd}$ and $3^\text{rd}$ peaks. 
%On the other hand, neural net assisted simulations show a fluid-like phase at the same temperature, with a weaker first peak and no other distinct peaks. 
%We estimate that there might be a phase transition around $k_BT=0.5$ for cubes. 
%Around the phase transition point, the static structure of the system is very sensitive to errors in interaction forces and torques, thus leading to the larger mismatch that we see in the middle panel of Fig. \ref{fig:pcf_03_cube}. 
%At lower and higher temperatures (i.e. temperatures that are further away from the phase transition point of the system) the prediction error introduced by the neural net predictions of the interactions seems to effect the $g(r)$ less. 

\begin{figure}
    \centering
    \includegraphics[width=8cm]{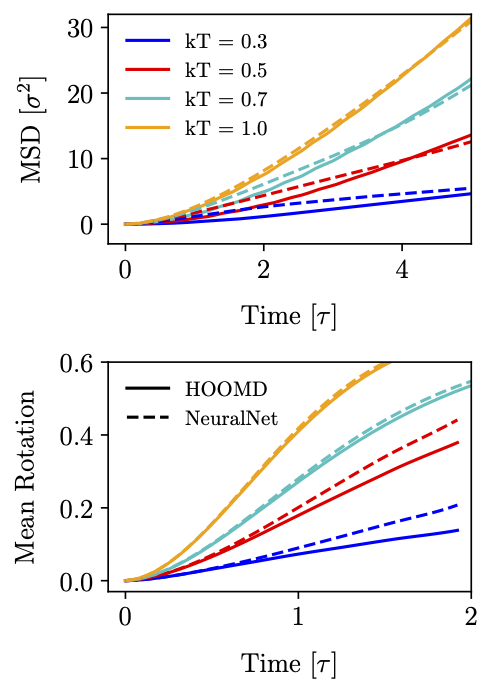}
    \caption{Mean squared displacement (top) and mean squared rotation (bottom) of cubes at different temperatures. The solid lines represent the traditional simulations, whereas the dashed lines show the nn-assisted results.}
    \label{fig:msr_cube}
\end{figure}

We also assessed the accuracy of dynamics in the neural-net assisted simulations. 
For consistency, the mass of the cube rigid-body is set to be the same in HOOMD and neural-net assisted simulations. 
In Fig. \ref{fig:msr_cube}, mean square displacements (MSD) of cubes at different temperatures for both neural-net assisted and traditional simulations are shown. At all temperatures, good agreement can be found. We did not determine long-term diffusion constants. Due to the matching in relative masses, we expect that the long term center of mass motion of the cubes will be similar in both  simulation techniques. In addition to  MSD, mean square rotation (MSR) is also a fundamental measure of the kinetics for non-spherical particles. 
MSR is the orientational-angular analogue of the MSD, and its calculation is explained in detail in the SI Section 5. 
In Fig. \ref{fig:msr_cube} the MSRs of the cubes at different temperatures are plotted. 
We observe a reasonable match, except at lower temperatures (0.5 and 0.3).  
Cubes in the nn-assisted simulations are rotating slightly faster compared to traditional MD simulations. %, however from $g(r)$ curves we also know that at that temperature traditional md simulation system is more ordered and aggregated and one would expect that the cubes in that system should rotate slower compared to the nn-assisted system where the cubes are still in fluid phase. 
It is not clear why we observe this behaviour, and future investigations will be needed.

Finally, the speed, or performance, of the neural-net assisted simulations were compared with HOOMD-blue simulation performance. 
Depending on the hardware used and on the number $N$ of cubes simulated, nn-assisted simulations can significantly increase the performance of simulations, as shown in Table \ref{tab:performance}.
We performed all simulations on a single RTX 2070 Super GPU with 8GB memory and a single A100 GPU with 80GB memory and report the timesteps per second.  An acceleration factor between the two simulation methods on each GPU was computed. Aside from the smallest system ($N=500$) on the A100, all nn-assisted simulations are faster than their corresponding HOOMD-blue simulations. In addition, on the RTX 2070, HOOMD rigid-body simulations run out of the 8GB memory for systems larger than $N\approx1300$.

\begin{table}
 \caption{Timesteps per second and acceleration factor of nn-assisted simulations compared to HOOMD-blue for various system sizes $N$. A dash (---) indicates that GPU memory was not sufficient to run the simulation.}
\begin{tabular}{l||*{6}{c}}
\backslashbox{GPU - Method}{N}
&\makebox[2em]{500}&\makebox[2em]{1000}&\makebox[2em]{1500}&\makebox[2em]{2000}&\makebox[2em]{3000}&\makebox[2em]{4000}\\[0.5em]\hline\hline
RTX-2070 - HOOMD & 9.5 &4.6 & --- & ---&---&---\\\hline
RTX-2070 - NN-Assisted &143 &109 & 90 & 75 & 64 & 45\\\hline
RTX-2070 - Speed up &\textbf{x15} &\textbf{x23} & --- & --- & --- & ---\\[0.5em]\hline
A100 - HOOMD &190 & 90& 60 & 45 & 30 & 23\\\hline
A100 - NN-Assisted &153 & 125& 108 & 91& 78 & 60\\\hline
A100 - Speed up & \textbf{x0.8} & \textbf{x1.4}& \textbf{x1.8} & \textbf{x2.0}& \textbf{x2.6}& \textbf{x2.6}\\[0.5em]
\end{tabular}
\label{tab:performance}
\end{table}

\section{Conclusions} 
In this work, we present a flexible neural-net assisted method to accelerate molecular dynamics simulations of non-spherical composite rigid bodies of any irregular shape. 
Here, we have used cubes and cylinders as examples, however, the method has no restrictions on the shape of the particles as long as sufficient training data can be obtained. Our method also allows any pair interactions, including repulsive and attractive forces. We observed excellent agreement in structural and dynamic properties between the nn-assisted and traditional rigid-body composite simulations. %, with some deviations close to a phase transition in the cubic system. 
We tested the performance of the current method and observed a speedup of x15 to x23 on a single RTX 2070 Super Nvidia GPU, and a speedup of x0.8 to x2.6 on a A100 Nvidia GPU. Depending on the system of interest and available hardware, this speedup can be significant.  

Our method has several shortcomings that prevent further speed-up, which can be addressed in future work. 
Ideally, the descriptors (for our model these are the input parameters of the neural-net) should be easy to calculate from the raw representation of the system, i.e., positions and quaternions~\cite{campos2022machine}.
However, the geometric operations that we perform to consider the symmetries of a pair of cubes or cylinders are computationally quite costly, reducing the performance. 
A second disadvantage is the use of a neural-net, where inference is associated with significant computational cost, especially considering inference is needed at every single timestep.  

We note that cylinders have an advantage where any pair configuration can be expressed with four numbers instead of six due to the cylindrical symmetry.
This symmetry greatly improved the performance of the neural-nets, or equivalently, reduces the difficulty of fitting. However, this symmetry assumption led to a slight loss in accuracy, particularly of parallel cylinder configurations at short distances. 
Depending on the desired density and level of surface roughness (i.e. how coarse-grained the cylinder is) these inaccurate interaction predictions might be significant.

% atomistic is easier, can include ligands, charges, depletion to this method
Our method, as presented here, is completely flexible in the input training data. Consequently, it can be easily extended to small nanoparticle systems that include ligands, charges, atomistic interactions, and other effects.
Generally, as particle sizes get smaller, the error in effective interactions between particles increases, due to classical approximations and coarse-graining \cite{silvera2015nonadditivity}.
We speculate that our approach could potentially attain the level of accuracy comparable to atomistic simulations by training the neural nets using nanoparticles made out of atoms.
Similarly, for small nanoparticles, surface ligands play a significant role in self-assembly, and neural-nets can be trained with effective forces and torques between nanoparticles with explicit ligands\cite{giunta2023coarse,campos2022machine}. 
Self-assembly pathways of charged colloids are also highly dependent on morphology \cite{rosenberg2020self}. 
It would be possible to include coulombic interactions by adding charged beads to the rigid bodies, which would be learned by the neural nets as a part of their effective interactions. 
Depletion interactions of nanoparticles depend strongly on their shape, and this method could provide an efficient way of developing effective depletion pair interactions by using traditional simulations with explicit depletants.

In this paper, the effective interaction we used was relatively short-ranged. It decayed to zero around $1.4a$, where $a$ is the size of the shape, which is shorter than a typical Lennard-Jones interaction that would decay to zero around $2.5$ to $3a$. 
Especially for nanoparticles, relatively long-ranged interactions can be relevant in self-assembly~\cite{kim2017imaging}.
Longer-range shape interactions would significantly increase the potential speedup of the nn-assisted method over traditional simulations. Unfortunately, more data sampling and possibly longer training times would be required.  
Similarly, more irregular shapes may require more composite beads for an accurate representation, which would lead to a larger potential benefit for the nn-assisted method. However, the regression task would be more complicated as well. 
Future research and methods development will push the limits of the method shown here, and we hope to include different physical interactions, longer-ranged interactions, and more complicated geometries. 

%%% upload the code 
%%% nanoscale horizons, jctc, JCP -> ALL related papers are here , Muller-Plathe Advanced THeory and Simulations 

\section*{Acknowledgements}
%\begin{acknowledgments}

This work was supported by funding from the Molecule Maker Lab Institute (MMLI): An AI Research Institutes program supported by NSF under award No. 2019897. This work made use of the Illinois Campus Cluster, a computing resource that was operated by the Illinois Campus Cluster Program (ICCP) in conjunction with the National Center for Supercomputing Applications (NCSA) and which was supported by funds from the University of Illinois at Urbana-Champaign.
%\end{acknowledgments}

\section*{Supplemental Information}
The supplementary information contains the details of the geometric operations performed to take the symmetries of pairs of cubes and cylinders into account. A discussion of the cylindrical symmetry assumption for cylinders is included. An example to demonstrate why many beads can be necessary for a rigid body is also included. We show how one can obtain torques and forces from the energy neural net. The code for running the nn-assisted simulations is available at \url{https://github.com/stattlab/NNAssistedRigidBodyMD}.

\bibliography{references}

\end{document}

\title{Supplementary Information: Molecular Dynamics Simulations of Anisotropic Particles Accelerated by Neural-Net Predicted Interactions}  % Adjust the title for supplementary material

\author{B. Ru\c{s}en Argun}
\affiliation{Mechanical Engineering, 
                              Grainger College of Engineering, University of Illinois,Urbana-Champaign, 61801, IL}

\author{Yu Fu}
\affiliation{Physics, 
                              Grainger College of Engineering, University of Illinois,Urbana-Champaign, 61801, IL}
                
\author{Antonia Statt}
\email{statt@illinois.edu}
\affiliation{Materials Science and Engineering, 
                              Grainger College of Engineering, University of Illinois, Urbana-Champaign, 61801, IL}

\date{\today}

\maketitle
\section{Data Sampling and Processing for Cubes } \label{sec:Data Sampling and Processing (Cubes) }
\begin{figure}[ht]
    \centering
    \includegraphics[width=0.49\textwidth]{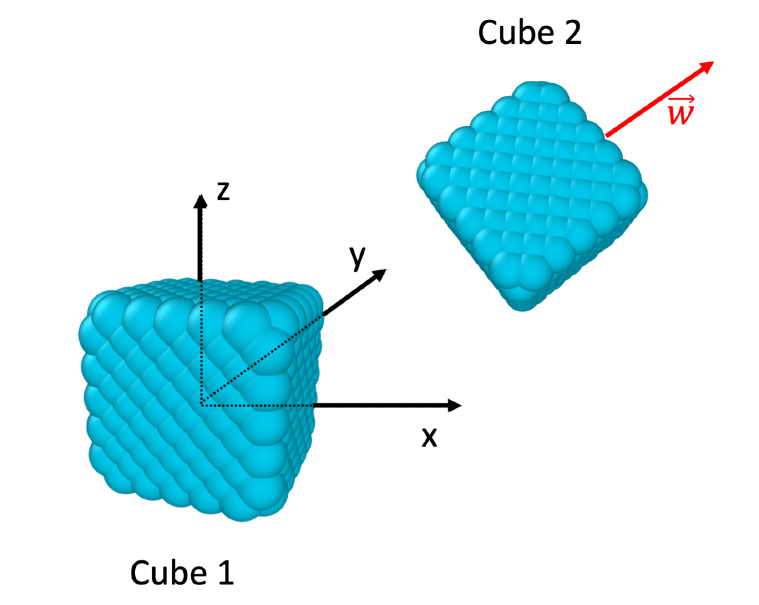}
    \caption{(a) Illustration of the relative coordinate frame of two cubes. $\boldsymbol{w}$ corresponds to the new orientation vector of cube 2 relative to cube 1. }
    \label{fig:cube_relative_coordinate_system}
\end{figure}

Pair configurations of cubes were sampled from HOOMD simulations that were run at different temperatures in the $0.2-2.0$ $k_BT$ range. A raw pair configuration sampled directly from the simulation is expressed by the following:
\begin{itemize}
\item Center of mass of cube 1: $\boldsymbol{p}_{\mathbf{1}}\left(p_1^x, p_1^y, p_1^z\right)$
\item Center of mass of cube 2: $\boldsymbol{p}_2\left(p_2^x, p_2^y, p_2^z\right)$
\item Orientation of cube 1: $\boldsymbol{q}_1\left(q_1^s, q_1^i, q_1^j, q_1^k\right)$
\item Orientation of cube $2: \boldsymbol{q}_2\left(q_2^s, q_2^i, q_2^j, q_2^k\right)$
\end{itemize}
Currently, the pair configuration is given by $3+3+4+4=14$ dimensions. This can be reduced to 6 by constructing a new coordinate frame that is centered at $\boldsymbol{p}_{\mathbf{1}}$ and oriented such that $\boldsymbol{q}_{\mathbf{1}}$ is $(1,0,0,0)$. For the cubes we are using, this means to have the coordinate axis $(i, j, k)$ aligned with the normal face vectors of the cube 1. There are 24 choices for picking the $(i,j,k)$ axis since the cube has 6 faces. We pick the axis directions such that $p_2^x, p_2^y$ and $p_2^z$ are all non-negative in this new coordinate frame, as illustrated in Fig.~\ref{fig:cube_relative_coordinate_system}.

\begin{figure}[ht]
    \centering
    \includegraphics[width=0.49\textwidth]{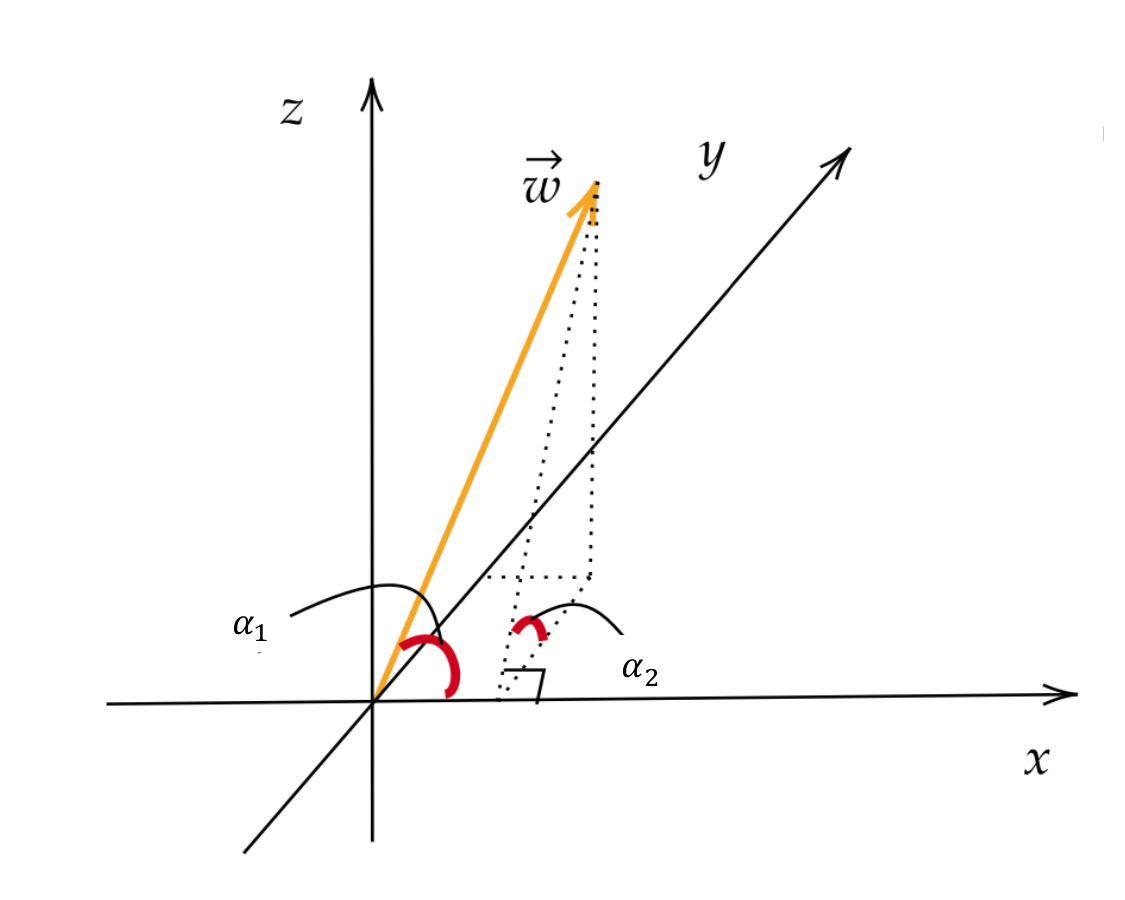}
     \includegraphics[width=0.49\textwidth]{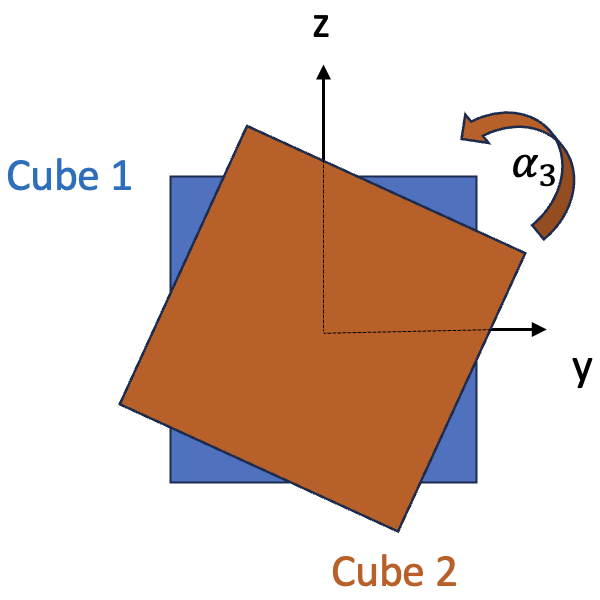}
    \caption{(a) $\alpha_{1}=\arccos(w^x)$ and $\alpha_{2}=\arctan2(w^z/w^y)$ are the $4^{\text {th }}$ and $5^{\text {th }}$ descriptors, respectively. They define the orientation of the cube 2's interacting face normal relative to cube 1's interacting face normal (always $(1,0,0)$) (b) $\alpha_{3}$ is the $6^{\text {th }}$ and last descriptor. Once the interacting face normals of the cubes are aligned  on $i$ the remaining degree of freedom is the rotation of cube 2 around it. }
    \label{fig:relative_orientation}
\end{figure}

In this new configuration, cube 2 can be reflected across the $y=z$ plane to obtain the same pair configuration, i.e., a pair with the same interaction energy, same force and torque component magnitudes. In order to avoid sampling the same configuration twice with different descriptors, we make sure that $p_2^y \geq p_2^z$. If $p_2^z>p_2^y$, we reflect the cube 2 across the $y=z$ plane.

After these steps, $\boldsymbol{p}_2\left(p_2^x, p_2^y, p_2^z\right)$ can be used as the first 3 components (descriptors) of the pair configuration. The remaining 3 numbers will be provided by the orientation of cube 2. First, cube 2 is translated to the origin. Then, we pick the face normal of cube 2 that is opposite of the closest face to the origin $(\vec{w})$ as can be seen in Fig.~\ref{fig:cube_relative_coordinate_system}. This vector can be aligned with the $(1,0,0)$ direction by two rotations. These two angles are the $4^{\text {th }}$ and $5^{\text {th }}$ numbers of the pair configuration as shown in Fig. \ref{fig:relative_orientation}a. After the normal vectors of cube 1 and 2 are aligned, there is only one degree of freedom left, the cube 2 can be rotated around $(1,0,0)$ as well. The amount of rotation that will make it overlap completely with cube 1 is the last descriptor of the pair configuration (Fig.\ref{fig:relative_orientation}b).

\begin{figure}[ht]
    \centering
    \includegraphics[width=0.49\textwidth]{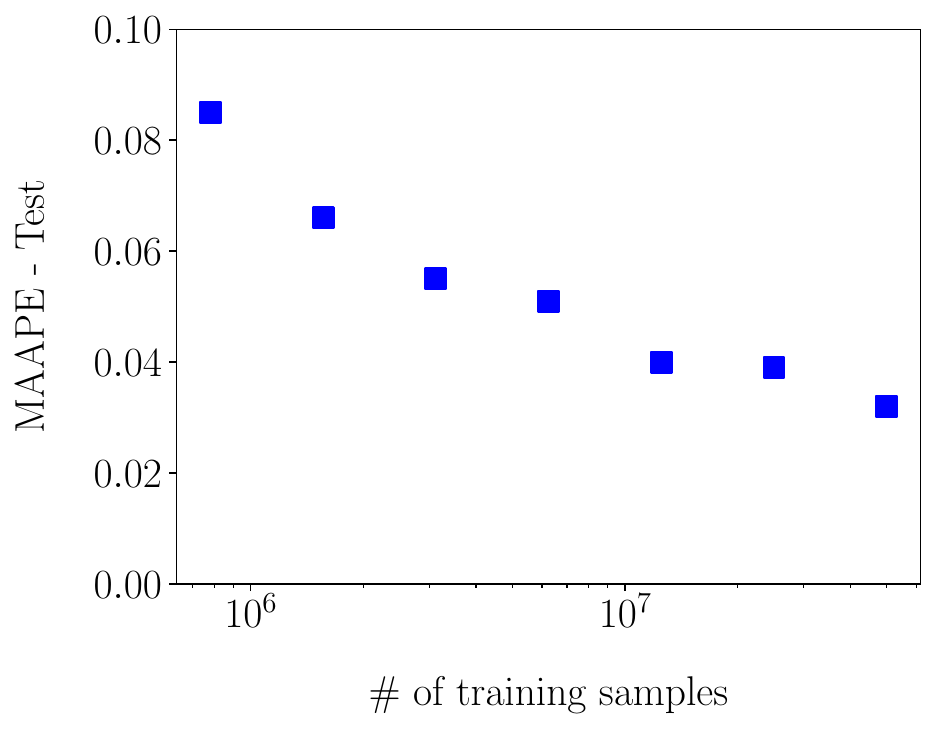}
    \caption{Test error rate vs. training data size }
    \label{fig:dsize_vs_error}
\end{figure}

We have used fully connected feed-forward artificial neural networks. Among the network sizes working well, the largest one had a width of 130 and depth of 15 (deeper nets seem to fail), smallest one had a width of 60 and a depth of 6. The test error slightly increases as the neural network size gets smaller, but the effect on the accuracy of the simulation was negligible. Therefore, considering computational cost, we have picked the smallest neural-network. The activation function is ReLU. We have used $5 \times 10^ 7$ samples to train the neural networks. 
We have also trained neural-networks with less amount of data and achieved similar error rates as can be seen in Fig.\ref{fig:dsize_vs_error}, where we show the Mean Arctangent Absolute Percentage Error (MAAPE) in energy prediction for cubes ~\cite{kim2016new} as function of training data set size. 

However, we have not performed any simulations with those neural-networks using smaller training sets. Therefore, we do not know the lower limits of the data required without significant simulation accuracy penalties. In the future,  simulation accuracy (i.e., not MAAPE errors) \textit{and} simulation performance (i.e., timesteps per second) should be optimized. 

\section{Data Processing for Cylinders} \label{sec:Data Processing (Cylinders)}
\begin{figure}[ht]
    \centering
    \includegraphics[width=0.5\textwidth]{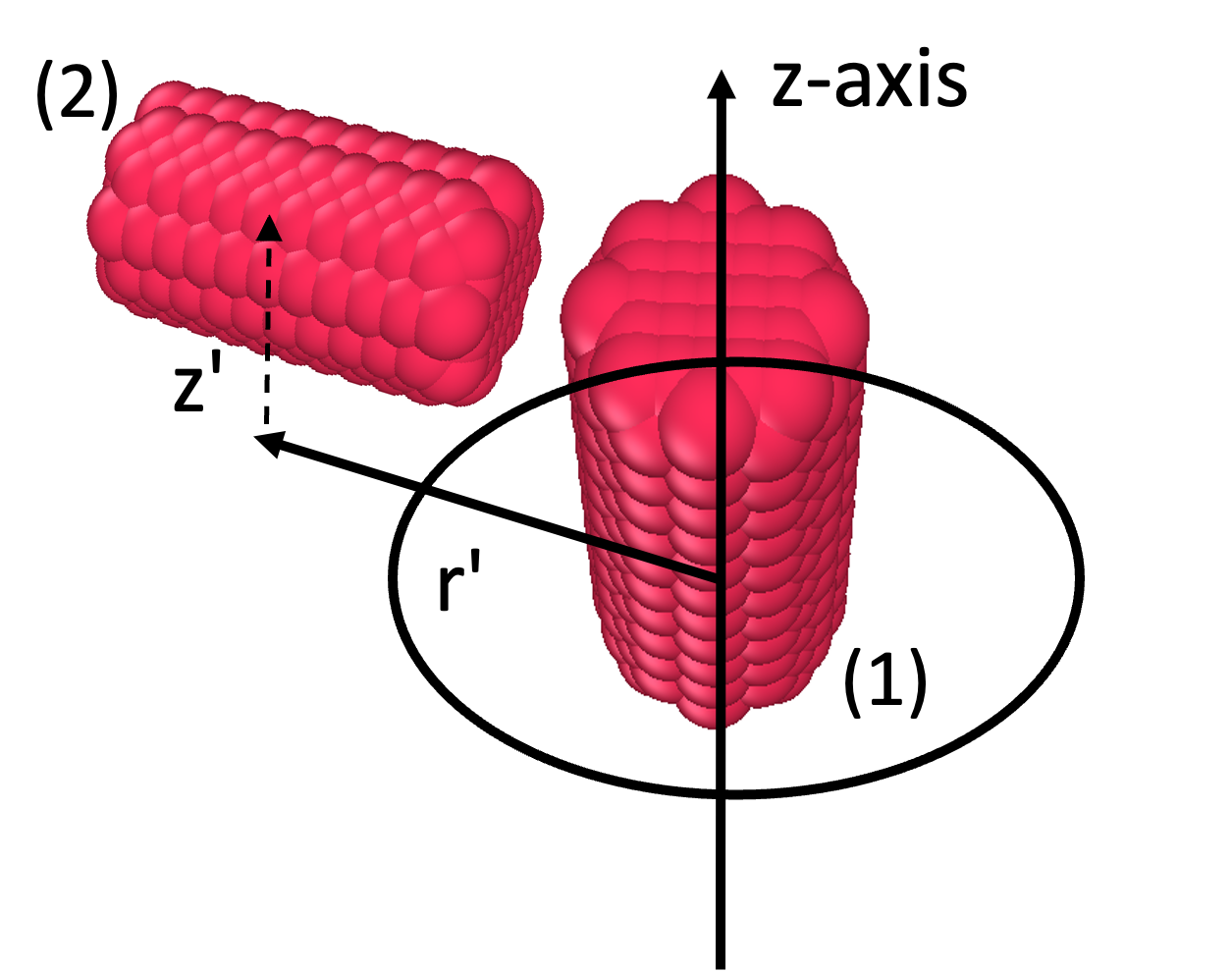}
    \caption{Illustration of  the relative coordinate frame for cylinders. }
    \label{fig:cylinder_coordinate_frame}
\end{figure}

Here, a similar process as for the cubes is described for the cylinders. One of the cylinders is picked as cylinder 1, and a new coordinate frame is constructed with its origin at the center of mass of  cylinder 1. The $z$ axis aligned with axis of cylinder 1, as shown in Fig.~\ref{fig:cylinder_coordinate_frame}. There are two choices for the direction of the $z$ axis. We pick the one direction such that $p_2^z$ is non-negative, to avoid sampling the same configuration twice. Then, we pick the $x$ direction of the coordinate frame such that $p_2^y=0$ and $p_2^x>0$. At this step, $\left(p_2^x, p_2^z\right)$ or effectively $\left(p_2^r, p_2^z\right)$ fully describes the position of the cylinder 2 with respect to to cylinder 1. Note that a third number is redundant, due to the assumed cylindrical symmetry of cylinder 1.

To find the relative orientation of cylinder 2, we consider the unit vector of its axis. This unit vector is converted into spherical coordinates and the two angles  are used as the other two descriptors for the pair configuration.

\section{Cylindrical Symmetry Assumption} \label{sec:Cylindrical Symmetry Assumption}
\begin{figure}[ht]
    \centering
    \includegraphics[width=0.8\textwidth]{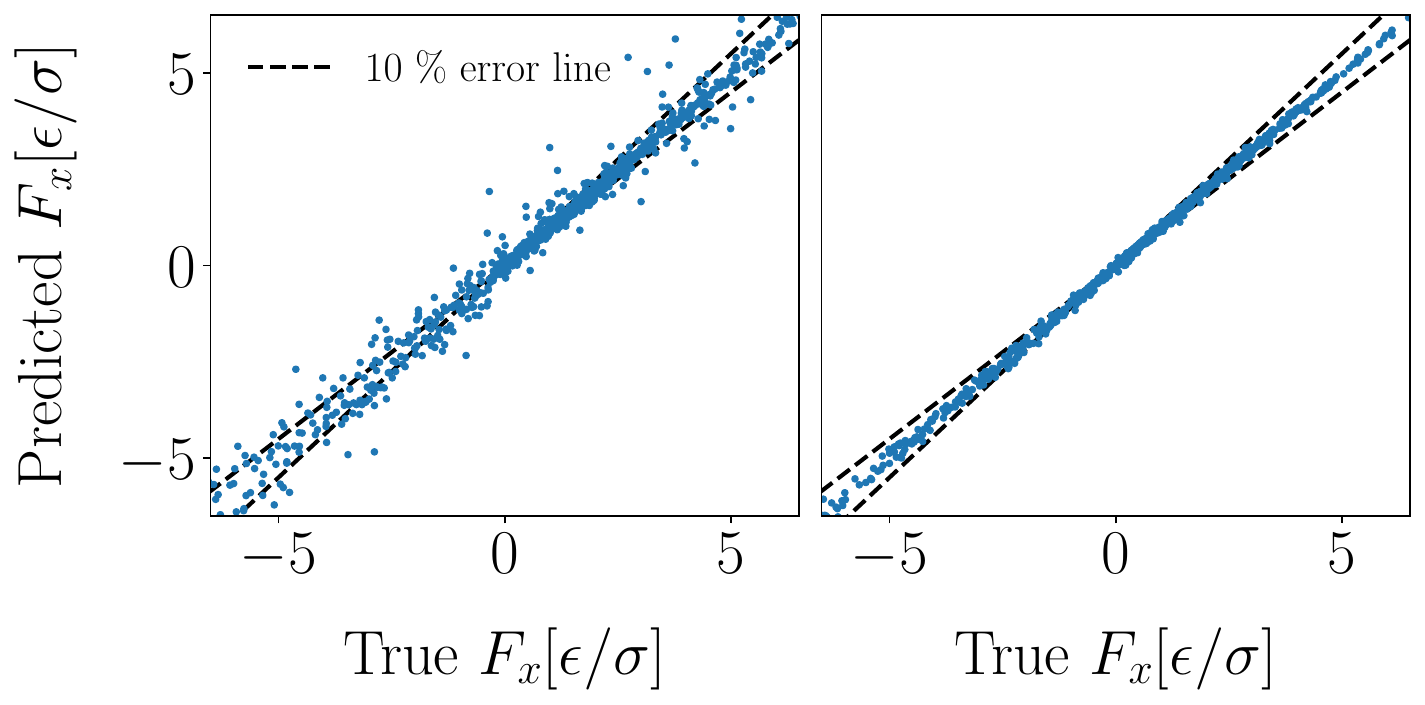}
    \caption{Parity plot of true vs. predicted $f_{x}$ values for the cylinder (left) and cube (right) system. Unlike the main text, for both cylinder and the cube, $f_{x}$ is the direct output of the force neural net $g_{NN_{f_{i}}}$. Dashed lines indicate the 10\% error line.}
    \label{fig:force_parity_SI}
\end{figure}

Any pair of $3D$ shapes in space can be described as a function of six numbers total. Thanks to their symmetry, a pair of cylinders can be described as a function of four numbers only. Considering our approach of predicting the interaction energy, force, and torque from the pair configuration, cylindrical symmetry significantly reduces the input dimension for the neural networks and thus simplifies the regression task considerably. Therefore, we expected to achieve similar accuracy when compared to cubes by using a smaller training data set for the cylinder system. However, this is not the case. In fact, the results for the cylinders are less accurate in structure, as seen in the pair correlation function $g(r)$. Additionally, there is a larger spread in the force parity plots, and the overall accuracy is lower as plotted in Fig.~\ref{fig:force_parity_SI}. This is mainly due to the composite-bead nature of the cylinders we simulated in traditional MD using HOOMD. Even though these composite shapes contain many beads, they are still not perfectly symmetric, the rugged surface of the cylinders breaks the assumed symmetry and causes errors in neural network predictions.  

\begin{figure}[ht]
    \parbox{0cm}{\includegraphics[width=0.5\textwidth]{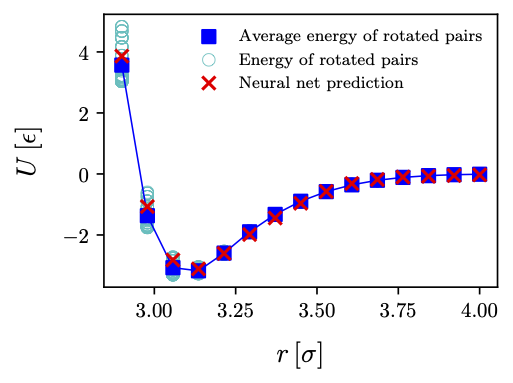}}
    \hspace*{9cm}
    \parbox{8cm}{\includegraphics[width=0.4\textwidth]{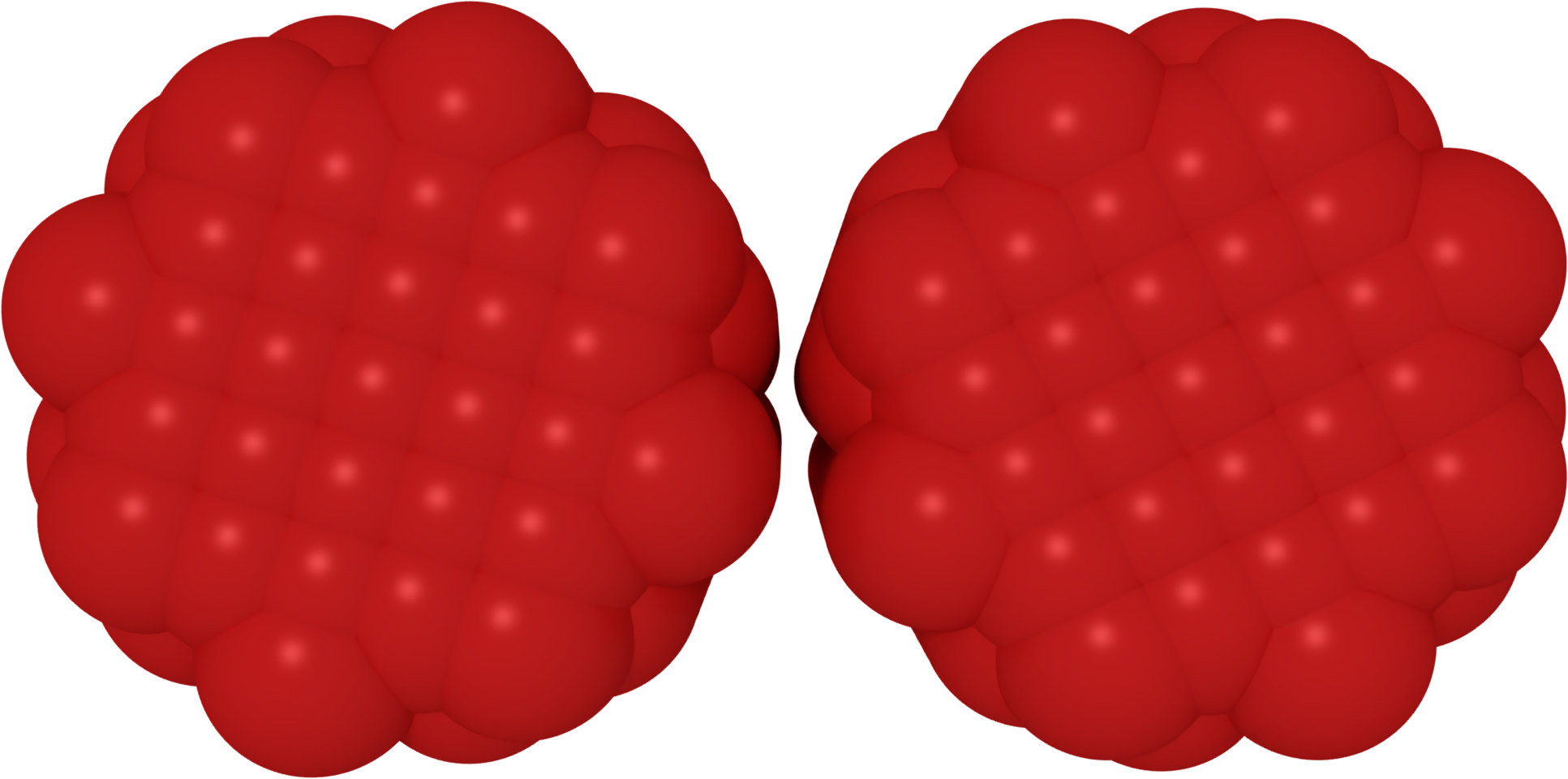}}
    \caption{(a) Energy vs. center of mass distance when cylinders’ axes are parallel. Cylinders are rotated around their axis incrementally at each center of mass distance. (b) A pair of aligned cylinders seen from top. Due to the discrete nature of the rigid bodies, the surface is rough.}
    \label{fig:uncert}
\end{figure}

This effect is illustrated in Fig.~\ref{fig:uncert}(a), where, at each distance separating two parallel cylinders, the neural-net predicted energy is compared to  the average energy as calculated from the composite cylinders by rotating them around their $z$ axis. While the NN result and the average agree reasonably well, there is a large spread in the composite shape energies. This is especially pronounced at small $r<3.2\sigma$ distances, where composite cylinders may ``interlock'' as can be seen in Fig. ~\ref{fig:uncert}(b). 
Overall, this illustrates that the neural net learned the \textit{average} interaction energy, and cannot faithfully represent the rugged surface of the composite shapes. Depending on the shape of interest, this behavior can be actually desired to average out some of the composite bead effects.

\section{Many beads requirement illustration } \label{sec:Many beads requirement illustration }
For a systematic study of the effect of the colloid shape on self-assembly, rheological behaviour etc. we may need to simulate systems of colloids with different shapes, or systems with mixtures of particle shapes. Assuming that we want to perform a generic comparison of spherical and cubic colloids or nanoparticles, van der Waals attractions must be considered. For simplicity, we assume that the spherical colloids interact with a Lennard-Jones potential, representing van der Waals-type interactions. The initial problem one encounters then is how to coarse-grain (i.e., tesselate with smaller composite beads) the spherical \textit{and} cubic colloids consistently, since that choice will have a strong effect on the effective interaction between the cube particles and how transferable parameters are between different shapes. 

To illustrate, we compare two different coarse-graining approaches and how transferable bead-to-bead interactions are in each case. The first one attempts to minimize the number of small composite beads, by placing them only on the surface of the shape, to reduce the computational cost of the simulation. Now, the composite bead-to-bead potential needs to be adjusted so that the effective (summed up) potential between the two spherical particles is as close as possible to the original desired potential, in this case a Lennard-Jones-like interaction. For demonstration purposes, we are using a half-power half cosine as a base function for the bead-to-bead potential. This function has a generic shape of a pair potential with short range repulsion, intermediate range attractions that eventually decay to zero at long distances. Given by the following functional form
\begin{equation}
\mathrm{U}(r)= \begin{cases}-A \cos (c(r-b)) & r>b \\ (b-r)^p-2 A & r \leq b\end{cases}\quad,
\end{equation}
it has four parameters that somewhat independently adjust the steepness of repulsive part ($p$), depth of the minimum ($A$), position of the minimum ($b$), and the decay rate of the attraction ($c$) i.e., range of the potential.

\begin{figure}[ht]
    \centering
    \includegraphics[width=0.5\textwidth]{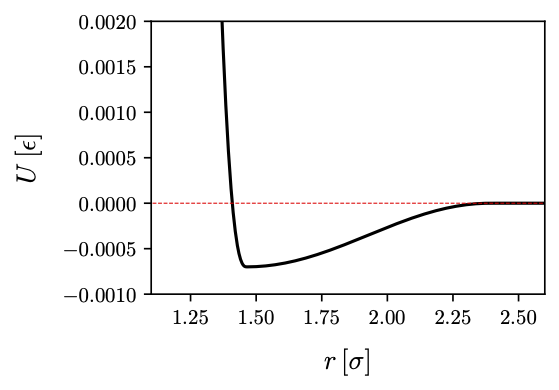}
    \caption{Half-power half cosine potential  }
    \label{fig:hphc_plot}
\end{figure}

\begin{figure}[ht]
    \centering
    \includegraphics[width=0.9\textwidth]{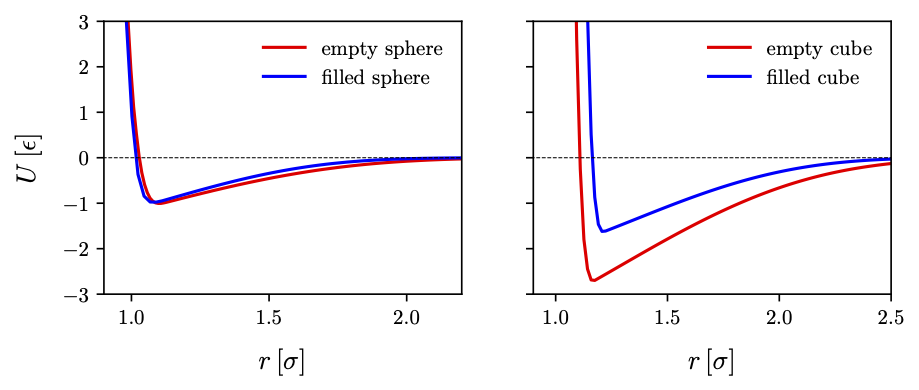}
    \caption{Effective interactions between a pair of empty and filled spheres (left) and effective interactions between a pair of empty and filled cubes (right) }
    \label{fig:best_fits}
\end{figure}

In Fig. \ref{fig:best_fits} (left - red curve) the best fit of individual composite bead parameters to the effective LJ function is plotted. Then, we use the exact same set of parameters($A, b, c$ and $p$) to calculate and plot the effective interaction between two cubes with opposite faces in \ref{fig:best_fits} (right - red curve). As expected, the position of the minimum is further away compared to the spherical particle due to the geometry of the cube. The attraction is stronger, since the flat faces of the cubes increase the maximum attraction.

% \begin{figure}[ht]
%     \centering
%     \includegraphics[width=0.8\textwidth]{SIfig/yellowbeads.png}
%     \caption{Two approaches can be used when coarse-graining rigid bodies. We can minimize the number of beads by placing them only to the surface  or fill inside as well like most real colloids are.}
%     \label{fig:coarse-graining_empty}
% \end{figure}

For the second approach we are using a sphere and a cube that are entirely filled with spherical beads in the inside as well, as most colloids would be. This will inevitably increase the number of beads needed, thus increasing the computational cost. We repeat the same process of fitting to a LJ potential with the spherical particle, the fits are shown in Fig. \ref{fig:best_fits} (left - blue curve). Notice that the fit to LJ is close to the fit obtained in the first approach for spherical particles. However, most importantly, when comparing the cubes for both approaches, the effective interactions are quite different. The depth of the attraction is 2.7 $\epsilon$ for the first coarse-graining method (empty particles) compared to 1.6 $\epsilon$ in the second one (filled particles). Meaning, that when considering transferability of interaction parameters between shapes, the choices made when tesselating the shapes with smaller beads are quite important. In this example, we know that van der Waals interactions scale with volume, the second approach would be more accurate and using only surface beads will result a different physical behavior. Depending on the system of interest and desired interactions, these considerations might be more or less important. For hard-core repulsion interactions, it would be sufficient to use surface beads only. 
\section{Mean Square Rotation Calculation } \label{sec:Mean Square Rotation Calculation }
The calculation follows directly from ~\citep{quat_distance}. The ``distance'' $d$ between two orientations can be given by the following:
$$
d\left(\boldsymbol{q}_1, \boldsymbol{q}_2\right)=1.0-\left\langle\boldsymbol{q}_1, \boldsymbol{q}_2\right\rangle^2
$$

Where $\boldsymbol{q}$ is the unit quaternion that describes the orientation of the particle. The inner product of two unit quaternions is given by the following:
$$
\left\langle\boldsymbol{q}_1, \boldsymbol{q}_2\right\rangle=\left\langle q_1^s+q_1^i \boldsymbol{i}+q_1^j \boldsymbol{j}+q_1^k \boldsymbol{k}, q_2^s+q_2^i \boldsymbol{i}+q_2^j \boldsymbol{j}+q_2^k \boldsymbol{k}\right\rangle=q_1^s q_2^s+q_1^i q_2^i+q_1^j q_2^j+q_1^k q_2^k
$$

After calculating this ``orientational distance'' between the starting orientation and orientations at each future timestep of the same particle, we average the distances of particles in the box to get the mean squared orientation at different conditions. Apart from the calculation of the orientational distance measure, this is similar to the usual mean square displacement calculations.

\section{Getting Forces and Torques from energy Neural-Net } \label{sec:ft_from_NN }

Net force and torque vectors on rigid bodies are required to perform molecular dynamics simulations of rigid bodies, so training a neural-net that can predict only the energy is not sufficient. 
One alternative is to train six other neural networks in addition to the energy neural net. 
Each neural net can be trained to predict a single component of the force $\mathbf{f}(f_1,f_2,f_3)$ or torque $\boldsymbol{\tau}(\tau_1,\tau_2,\tau_3)$. 
We show that this approach works but has some disadvantages. 
First, training six separate neural-net requires extra computational effort. 
Also, the amount of data to be sampled and processed before the training is six-fold compared to a single energy neural net. 
Second, since force and torque are derivatives of the energy neural net and the range of possible values that we can have in the simulation is much wider compared to the energy. 
For example, all sampled energy values for pair of cubes were in range [-5.1,17.0] $k_BT$ whereas sampled $f_1$ values can range from [-10,120]. 
The wide range distribution of force and torque values poses an extra challenge for training and reduces the test accuracy so in order to reduce the error rate, we have capped the force and torque values in the training data, i.e. we've sacrificed the accuracy in extreme force and torque values for higher accuracy in values towards the middle of the range.  
However to train the energy neural-net, such measures and sacrifices are not necessary. 
So instead, it is more practical to use the energy neural network and calculate the forces by taking the gradients of the energy neural network. 

\subsection{Forces}
For a pair of cubes, with the way input parameters of the energy neural network are setup, it is relatively simple to obtain the forces. 
The first three inputs of the neural-net are already the relative position coordinates of cube 2 $(x_{1}, x_{2}, x_{3})$. 
Derivative of the energy neural net with respect to the position coordinates can be numerically calculated. 
\begin{equation*}
    f_{i} = -\frac{\partial E}{\partial x_{i}}  \approx -\frac{E(x_{i}+\Delta x) - E(x_{i}-\Delta x)}{2\Delta x} \quad,
\end{equation*}
where $E$ is the energy predicted by the neural-net. 
\subsection{Torques}
The torque~\cite{allen2017computer} around an axis $\eta$ is given by the following:
\begin{equation*}
    \tau_{\eta} = -\frac{\partial E}{\partial \theta_{\eta}} \quad.
\end{equation*}
where $\theta_{\eta}$ is the angle of rotation around the axis $\eta$.

Unfortunately, due to the way input parameters are set-up we cannot simply take the derivatives with respect to the input components of the neural-net since $\theta_{\eta}$ is not an input parameter but a function of input parameters. 
Chain-rule can be applied to relate derivatives with respect to the input parameters $\frac{\partial}{\partial x_{i}},\frac{\partial}{\partial \alpha_{i}}$ (which we can  calculate numerically) to the derivative with respect to rotation around axis $\frac{\partial}{\partial  \theta_{\eta}}$, which we need to obtain the torques.
It is natural to pick $x,y,z$ as the three axis for torque components.
Then the torque around $z$-axis, $\tau_{z}$ can be expressed as: 
\begin{equation*}
    \tau_{z} = -\frac{\partial E}{\partial \theta_{z}} \quad.
\end{equation*}

Enumerated out, that results in
\begin{align*}
    \tau_{z} &= -\frac{\partial E}{\partial \theta_{z}} = -\sum_{i=1}^{3}\frac{\partial E}{\partial x_{i}}\frac{\partial x_{i}}{\partial \theta_{z}} - \sum_{i=1}^{3} \frac{\partial E}{\partial \alpha_{i}}\frac{\partial \alpha_{i}}{\partial \theta_{z}}\\
    %\tau_{z} &= -\frac{\partial E}{\partial \theta_{z}} \\
    &= -\frac{\partial E}{\partial x_{1}}\frac{\partial x_{1}}{\partial \theta_{z}} - \frac{\partial E}{\partial x_{2}}\frac{\partial x_{2}}{\partial \theta_{z}} - \frac{\partial E}{\partial x_{3}}\frac{\partial x_{3}}{\partial \theta_{z}} \\
    &- \frac{\partial E}{\partial \alpha_{1}}\frac{\partial \alpha_{1}}{\partial \theta_{z}} - \frac{\partial E}{\partial \alpha_{2}}\frac{\partial \alpha_{2}}{\partial \theta_{z}} - \frac{\partial E}{\partial \alpha_{3}}\frac{\partial \alpha_{3}}{\partial \theta_{z}}\quad.
\end{align*}

Next, the goal is to express the unknown terms $(\frac{\partial x_{i}}{\partial \theta_{z}},\frac{\partial \alpha_{i}}{\partial \theta_{z}})$ in the equation above as a function of known terms i.e. input parameters of the neural net $(x_{i},\alpha_{i})$. 
It is more convenient work in cylindrical coordinates $(r,z,\theta)$ that is centered at cube 1, such that the $z$-axis around which we want to calculate the torque is the $z$-axis of the cylindrical coordinate and the rotation angle around that axis $\theta_{z}$ is equivalent to $\theta$ of the cylindrical coordinates. 
Using $x=r\cos(\theta)$, $y=r\sin(\theta)$ and $z_{cartesian}=z_{cylindrical}=z$, 
\begin{align*}
    \frac{\partial x_{1}}{\partial \theta_{z}} &= \frac{\partial x}{\partial \theta} = \frac{\partial r\cos(\theta)}{\partial \theta} = -r\sin(\theta) = -y = x_{2}\\
  \frac{\partial x_{2}}{\partial \theta_{z}} &= \frac{\partial y}{\partial \theta} = x = x_{1}\\
    \frac{\partial x_{3}}{\partial \theta_{z}} &= \frac{\partial z}{\partial \theta} = 0\quad.
\end{align*}

The last three terms involve the 3 angles $(\alpha_{1},\alpha_{2},\alpha_{3})$ that describe the orientation of cube 2. 
$\alpha_{1}$ and $\alpha_{2}$ are defined with $\mathbf{w}$ face normal unit vector of cube 2 as explained previously, $\alpha_{1}=\arccos(w^x)$ and $\alpha_{1}=\arctan2(\frac{w^z}{w^y})$. 
With the help of cylindrical coordinates we find
\begin{align*}
    \frac{\partial \alpha_{1}}{\partial \theta_{z}} &= \frac{\partial \arccos(w^x)}{\partial \theta_{z}} = \frac{\partial \arccos(r \cos(\theta))}{\partial \theta} = \cos(\alpha_{2}) \\
    \frac{\partial \alpha_{2}}{\partial \theta_{z}} &= \frac{\partial \arctan2(\frac{w^z}{w^y})}{\partial \theta_{z}} = \frac{\partial \arctan2(z/rsin(\theta))}{\partial \theta} =\frac{-\cos(\alpha_{1})\sin(\alpha_{2})}{\sin(\alpha_{1})} \\
    \frac{\partial \alpha_{3}}{\partial \theta_{z}} &=\frac{1-\cos(\alpha_{1})\sin(\alpha_{2})}{\sin(\alpha_{1})} \quad.
\end{align*}

The remaining equations that give $\tau_{x}$ and $\tau_{y}$ can be found in the code provided.

% \input{sections/acknowledgements.tex}

% The bibliography in supplementary material may also be handled differently.
% It might be integrated into the main article's bibliography or kept separate.
% \begin{thebibliography}{4}
% \bibitem{Griffiths}
% D. J. Griffiths,
% \textit{Introduction to Electrodynamics}
% (Cambridge University Press, Cambridge, 2017).

% \bibitem{Fleming}
% A. Bobrinha,
% Revista Brasileira de Lorem Ipsum \textbf{23},
% 179 (2002).

% \bibitem{Feynman}
% R. P. Feynman, R. B. Leighton and M. Sands,
% \textit{Lições de Física de Feynman}
% (Editora Bookman, Porto Alegre, 2008).

% \bibitem{Jackson-CE}
% J. D. Jackson,
% \textit{Classical Electrodynamics}
% (John Wiley \& Sons, Danvers, 1999).
% \end{thebibliography}

% \appendix*
% \input{sections/appendix1.tex}
\bibliography{SI_refs}